\documentclass[twocolumn,showpacs,preprintnumbers,amsmath,amssymb,superscriptaddress]{revtex4}
\usepackage{graphicx}
\usepackage{dcolumn}
\usepackage{bm}
\graphicspath{{./Figures/}}

\begin{document}

\preprint{APS/123-QED}

\title{Extended space charge near non-ideally selective membranes and nanochannels}

\author{Jarrod Schiffbauer}%
\author{Neta Leibowitz}%
\author{Gilad Yossifon}%
\affiliation{%
Faculty of Mechanical Engineering, Micro- and Nanofluidics Laboratory, Technion - Israel Institute
of Technology - Technion City 32000, Israel
}%

\date{\today}

\begin{abstract}
We demonstrate the role of selectivity variation in the structure of the non-equilibrium extended space-charge using 1D analytic and 2D numerical Poisson-Nernst-Planck models for the electro-diffusive transport of a symmetric electrolyte. This provides a deeper understanding of the underlying mechanism behind a previously-observed maximum in the resistance-voltage curve for a shallow micro-nanochannel interface device [Schiffbauer, Liel, Leibowitz, Park, and Yossifon,~\emph{submitted to Phys. Rev. E.}]. The current study helps to establish a connection between parameters such as the geometry and nanochannel surface-charge and the control of selectivity and resistance in the over-limiting current regime.
\end{abstract}

\maketitle
\section{Introduction}
\indent The current-voltage response of fabricated micro-nanochannel devices and ion-selective membranes are in many respects similar. Both exhibit ion-selective transport and produce concentration polarization (CP) in adjacent solution resulting in a near limiting saturation of the DC current response. At sufficiently high voltages, both exhibit an over-limiting current (OLC) which may be associated with electro-convective vortices~\cite{RZPRE2000,RZPRL2008,YCPRL2008,Eugene} among other mechanisms~\cite{DydekPRL,Yaroschuk,NBruus,KimPRL2015}. However, there are a number of differences between ion-selective membranes, such as Nafion, and fabricated nanochannel devices. The geometry of pores within a membrane are often irregularly shaped, highly tortuous, and poorly connected~\cite{NafionMemb}, and the fixed charge density quite high. One critical difference is that, while membranes such as Nafion typically exhibit near ideal selectivity across a wide range of parameters~\cite{NafionSelect}, nanochannel selectivity can vary appreciably. The similarity between membranes and nanochannels implies that a non-ideal membrane model can serve as a simple model for electro-diffusive transport through a nanochannel.\\
\indent Previous studies have shown variation in selectivity in the under-limiting regime~\cite{AbuRjal} and demonstrated the role of non-ideality in rectification due to diffusion-layer asymmetry~\cite{dissertation}. Other studies concern the role of a non-equilibrium extended space charge (ESC) on transport of competing counter-ions through an ideal membrane~\cite{Zabolotsky}, or consider space-charge dynamics for a binary electrolyte through a non-ideal membrane in the over-limiting current (OLC) regime~\cite{RZPRE2010}. However, these previous studies omit the first-order correction to the salt-concentration in the space charge regime and its coupling to the selectivity. Here we demonstrate the necessity of including these terms in capturing the effects of changing selectivity on the resistance maximum. This is expected to have consequences for the observability of the resistance maximum in real systems, where other OLC mechanisms can enhance or compete with the ESC driven resistance maximum~\cite{NewPRE_R}. Furthermore, it may have implications for the control of selectivity and ESC structure in practical applications, such as bio-molecular sensing, analyte pre-concentration, or electrokinetic desalination.\\
\section{Theory}
\begin{figure}
  \includegraphics[width=3.5 in]{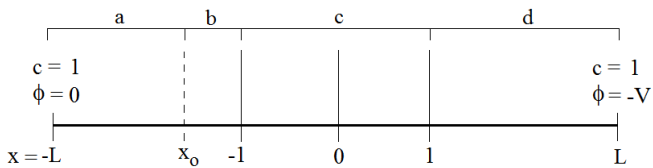}\\
  \caption{Basic geometry for the 1D model. Regions a and d are respectively the depleted and enriched quasi-electroneutral diffusion layers, region c is charge-selective, and region b approximates the width of the ESC.}\label{fig:f1}
\end{figure}

\begin{figure}
  \includegraphics[width=3.5 in]{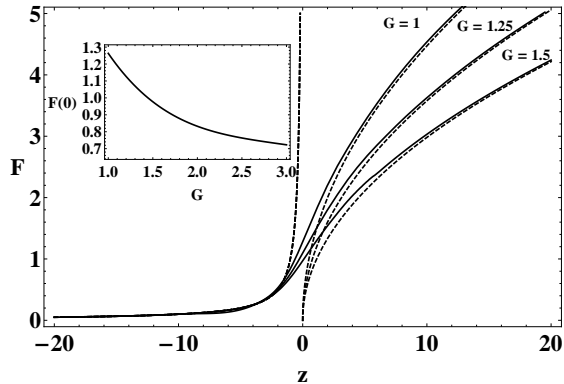}\\
  \caption{Numerical solution to modified Paineleve equation (Eqn.~\ref{eqn:e19}) for several values of $G$ (solid black). Large-z asymptotes (Eqn.~\ref{eqn:e20}) are shown also (dashed black.) The inset shows the interpolated value of $F(0)$ as a function of $G$, used in subsequent calculations.}\label{fig:f2}
\end{figure}

\begin{figure}
  \includegraphics[width=3.5 in]{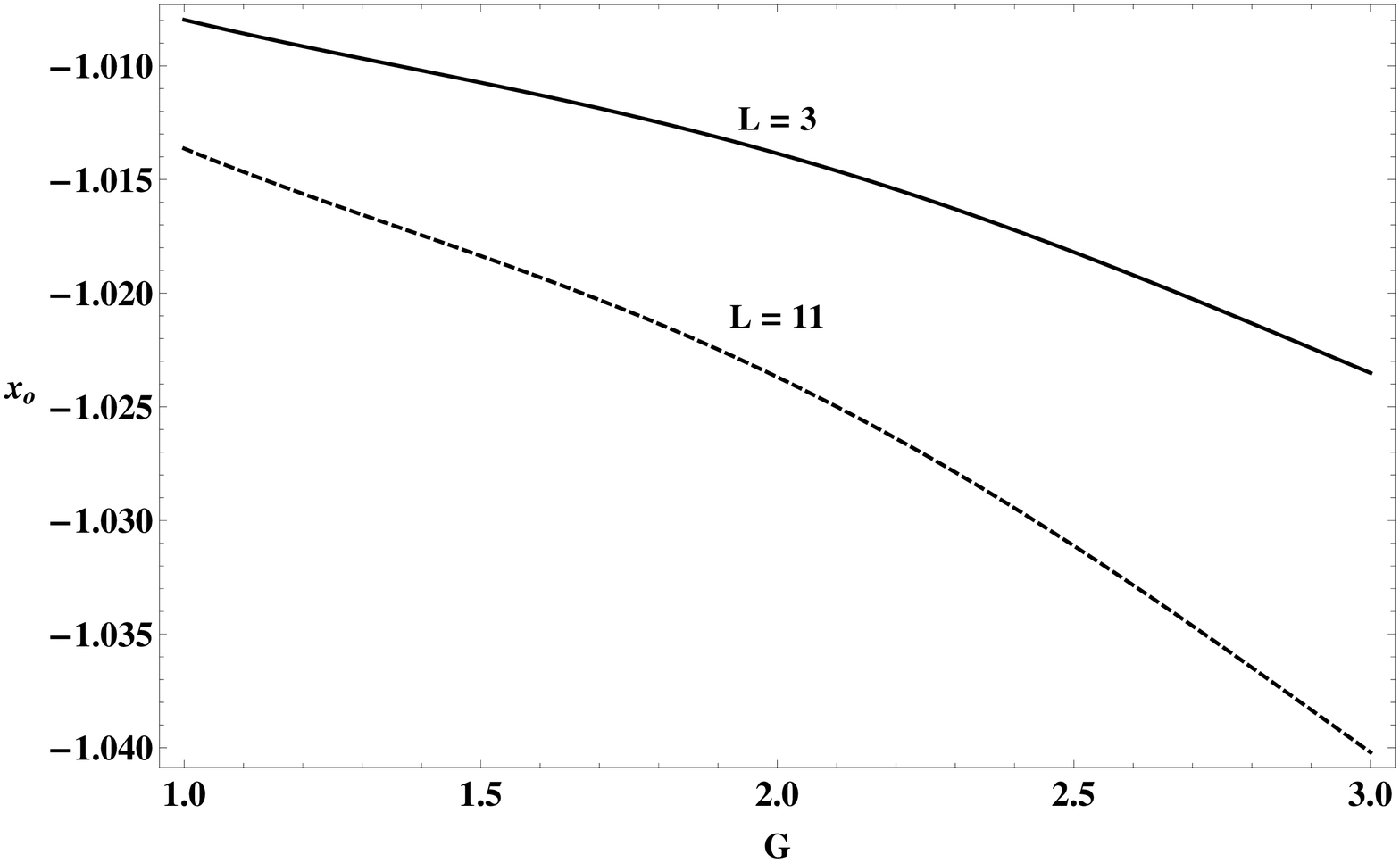}\\
  \caption{Estimate of the ESC width at the limiting salt flux density as a function of selectivity for $\epsilon=0.001$. Two cases are shown, for diffusion layer length equal to the membrane thickness and for a diffusion layer much thicker than the membrane. For thinner diffusion layers, the result is (unsurprisingly) that the ESC is thinner overall and the change in thickness is less.}\label{fig:f3}
\end{figure}

\indent In the following, we consider one-dimensional steady-state electro-diffusive transport of a symmetric, binary electrolyte through a system consisting of two  solution layers flanking a non-ideally cation-selective membrane (see Fig.~\ref{fig:f1}). We omit effects such as surface conduction and fluid-flow for simplicity. In particular, we focus our attention on the modulation of charge-selectivity in the over-limiting regime and its effect on the previously observed resistance maximum~\cite{NewPRE_R}. Using the steady-state (non-dimensional) Poisson-Nernst-Planck (PNP) equations as a starting point, the ion concentration field is given by,
\begin{equation}\mathcal{D}(x)\left[-\frac{d c_{\pm}}{d x}\mp c_{\pm}(x)\frac{d \phi}{dx}\right]=j_{\pm} \label{eqn:e1}\end{equation}
for each species $\pm$, and the Poisson equation describes the electric potential
\begin{equation}-\epsilon^2 \frac{d^2 \phi}{d x^2} =c_+(x)-c_-(x)-N(x)  \label{eqn:e2}\end{equation}
where the (negative) fixed charge density is $N(x)=N$ for $x\in\left\{-1,1\right\}$ and and $0$ elsewhere. Lengths have been scaled by the membrane half-thickness, $\ell$, concentration by the bulk concentration, $c_o$, fixed at the ends, and the electric potential by the thermal voltage, $kT/e = 0.0254$ V. The ion diffusivity is scaled by the bulk value so that $\mathcal{D}(x)=1$ in the electrolyte, but in general may have a different value, $\mathcal{D}_m$, in the membrane interior. The applied field, hence electric current, is directed towards the right (positive) so the ESC sub-layer forms at the $x=-1$ interface as indicated in Fig.~\ref{fig:f1}. The dimensionless Debye length, $\epsilon =\sqrt{\varepsilon\varepsilon_o kT/e^2 L c_o}$ is taken as a small parameter, $\epsilon\ll1$. Its appearance in Eqn.~\ref{eqn:e2} multiplying the highest derivative of the potential introduces the possibility of boundary layers and singular perturbation methods are typically used to obtain an analytic solution to the full problem~\cite{JFM2007}.\\
\indent The solutions to the 1D problem are well-known in various approximations. Here we review results relevant to the non-ideal case. The problem is re-cast in terms of $c=(c_+ + c_-)/2$ (salt concentration,) the salt-flux density $J_+=(j_++j_-)/2$, and charge flux density $J_-=(j_+-j_-)/2$. The two flux densities are related by the selectivity factor $G=J_-/J_+$ and the electric current density is $I=2J_-$. While the full solution requires either singular perturbation theory or numerical methods, a useful first approximation may be obtained by taking the leading-order solution, or equivalently, setting $\epsilon=0$, which is tantamount to imposing local electro-neutrality (LEN). These locally electro-neutral solutions for the salt concentration in the 1D problem are bi-linear functions of the salt-flux density and the position,
\begin{equation}c_l(x)=1-J_+(L+x),\label{eqn:e3}\end{equation}
in the depleted (left) side and
\begin{equation}c_r(x)=J_+(L-x)+1\label{eqn:e4}\end{equation} in the enriched (right) side. The potentials are given by the integrals over the inverse concentration, or
\begin{equation}\phi(x)=G\ln{\left[1-J_+(L+x)\right]} \label{eqn:e5}\end{equation}
for the depleted region and
\begin{equation}\phi(x)=G\ln{\left[J_+(L-x)+1\right]}-V\label{eqn:e6}\end{equation}
for the potential in the enriched region. The difficulty with the LEN approximation can be seen clearly in Eqns.~\ref{eqn:e3} and~\ref{eqn:e5}. As the current approaches a value corresponding to the salt-flux density $J_{+,lim}=1/(L-1)$, the salt concentration at the depleted interface, $x=-1$, approaches zero and the corresponding electric potential tends to infinity, i.e. it requires an infinite voltage drop to reach this value of current. However this current can not only be reached but exceeded in real systems. Depending on the details of the system~\cite{DydekPRL,KimPRL2015}, the mechanisms which drive OLC can depend on violations of local electro-neutrality via the formation of the ESC~\cite{RUBSHTILL79}.\\
\indent Inside the ion-selective region, $x\in\left\{-1,1\right\}$, we assume local quasi-electro-neutrality to hold to arbitrarily high currents. So Eqn.~\ref{eqn:e2} reduces to $c_+ = c_- + N$. The question of practical limits on this condition, especially in the context of the micro-nanochannel interface, is beyond the scope of the present study. The governing equations for the salt concentration and potential in the membrane interior are obtained from Eqns.~\ref{eqn:e1} and~\ref{eqn:e2},
\begin{equation}\frac{dc_m}{dx}+\frac{N}{2}\frac{d \phi_m}{dx}=-\frac{J_+}{\mathcal{D}_m} \label{eqn:e7}\end{equation}
and
\begin{equation}\frac{d \phi_m}{dx} =-\frac{J_-}{\mathcal{D}_m}c_m^{-1} \label{eqn:e8}.\end{equation}
These are integrated to yield the following transcendental equation for the concentration profile,
\begin{eqnarray}& \left[c_m (x) -c_m (-1)\right] + \frac{N}{2}G \ln{\left[\frac{c_m(x) -\frac{N G}{2}}{c_m(-1)-\frac{N G}{2}}\right]}= \nonumber \\ & -\frac{J_+}{\mathcal{D}_m}\left(x+1\right) \label{eqn:e9},\end{eqnarray}
and
\begin{equation}\phi_m(x)=-\frac{J_-}{\mathcal{D}_m}\mathfrak{F}(x) + \phi_m(-1) \label{eqn:e10}\end{equation}
for the electric potential inside the ion selective region, where the integral
\begin{equation} \mathfrak{F}(x)=\int^{x}_{-1}c^{-1}_m (x') dx'\end{equation}\label{eqn:e11}
is defined for convenience.\\
\indent For salt-flux densities lower than the limiting value, the response of the system may be obtained from the boundary conditions at $x=\pm L$ by assuming continuity of the electrochemical potentials across the quasi-equilibrium electric double layers (EDLs) between each region, as first proposed by Kirkwood~\cite{Kirkwood}. Strictly speaking, this is an assumption of local, homogenous thermodynamic equilibrium between two physically distinct phases, separated by a defining, infinitesimally thin Gibbs dividing plane. Because the EDLs are quite thin and the structure appears to maintain a quasi-equilibrium character under current, this approximation may be employed in non-equilibrium scenarios. The extent of the validity of the internal quasi-equilibrium structure has been examined thoroughly and shown to be maintained well into the non-equilibrium regime~\cite{JFM2007}. The system retains a quasi-equilibrium EDL as the innermost sub-layer, with the ESC developing as an adjacent structure~\cite{RZPRE2010}. Thus one should always be able to impose, at least approximately, a continuity condition to jump across this inner EDL boundary, resulting in a Donnan-like equilibrium between the interior of the ESC (adjacent to the thin EDL) and the membrane interior. We employ this assumption herein.\\
\indent For the full solution, we follow the usual procedure for obtaining a master equation for the scaled E-field, $E=-\epsilon\partial \phi/\partial x$ (see for instance Ref~\cite{JFM2007}.) The main results are the following equation for the scaled E-field,
\begin{equation}\epsilon^2 \frac{d^2 E}{dx^2}-\frac{E^3}{2}+2J_+ E (x-x_o)=-2\epsilon G J_+ \label{eqn:e12}\end{equation}
and the equation for the cation concentration,
\begin{equation}c_+(x) =\frac{\epsilon}{2}E'+\frac{E^2}{4}-J_+(x-x_o) \label{eqn:e13}\end{equation}
and the salt concentration,
\begin{equation}c(x)=\frac{E^2}{4}-J_{+}(x-x_o) \label{eqn:e14}\end{equation}
Thus the concentrations may be calculated once a solution for the E-field is obtained. Note the integration constant $x_o$. It can be associated with the intersection of the extrapolated depleted linear salt concentration profile to zero concentration. At and above the limiting current, this point moves into the region $x\in\left\{-L,-1\right\}$, yielding an estimate of the edge position of the ESC.\\
\indent Asymptotic expansions for the scaled E-field are given by the following for regions sufficiently far ($\gg \epsilon^{2/3}$) from either side of the point $x=x_o$:
\begin{eqnarray}
E(x) = \left\{
        \begin{array}{ll}
            \frac{-\epsilon G}{(x-x_o)} + \frac{\epsilon^3(4G-G^3)}{4J_+(x-x_o)^4} + O(\epsilon^5) & \nonumber \\
            \quad \quad \nonumber \\
            2\sqrt{J_+(x-x_o)}+\frac{\epsilon G}{2(x-x_O)} + O(\epsilon^2) &\\
        \end{array}
    \right. \\
\label{eqn:e15}\end{eqnarray}
The first expansion is valid for regions to the left of $x_o$, corresponding to the outer edge of the ESC and (at not too high a voltage) a quasi-LEN diffusion layer qualitatively similar to Eqn.~\ref{eqn:e3}. The second expansion is valid within the ESC itself and can be integrated to find the voltage drop across the ESC to leading order,
\begin{equation}\Delta V_{ESC} = - \frac{4\sqrt{J_+}}{3 \epsilon}\left(-1-x_o\right)^{3/2} \label{eqn:e16}\end{equation}
Note that while this gives a large ($O(\epsilon^{-1})$) contribution to the voltage drop, hence dominating the response in terms of voltage, the lowest-order non-vanishing contribution to the concentration in the ESC arises from the first-order correction. It was shown in~\cite{RZPRE2003} that this approach yields a well-defined minimum in the marginal stability of the ESC. Because this contribution affects the structure and resistance of the ESC and is also coupled to the selectivity of the membrane, we choose to employ it calculating the ESC-membrane Donnan condition.\\
\indent A rigorous treatment of the problem (see Ref.~\cite{JFM2007}) involves defining appropriate boundary layer variables and careful consideration of the behavior of the (transformed) parameter $x_o$ across the full range of applicable applied voltages in the limit $\epsilon \rightarrow 0$. Here we adopt a less rigorous approach, similar to that employed in~\cite{RZPRE2003}, to demonstrate the coupling between ESC structure, non-ideal selectivity, and the Ohmic-to-OLC transitional ESC resistance maximum~\cite{NewPRE_R}. The idea is akin to solution-patching with a first-order correction and demanding that Eqns.~\ref{eqn:e3} and~\ref{eqn:e15} yield the same value for the concentration at our estimated ESC edge, $x=x_o$.\\
\indent Transforming to the boundary layer variables, $E=(2J_+\epsilon)^{1/3}G F$ and $x-x_o=(2J_+)^{-1/3}\epsilon^{2/3}z$, the salt concentration at the ESC edge is given by
\begin{equation}c(x_o)= \frac{(\epsilon 2J_+)^{2/3}G^2 F^2(0)}{4} \label{eqn:e17}\end{equation}
so that the ESC edge position can be approximated by,
\begin{equation} x_o = \frac{1}{J_+}-L-(2\epsilon)^{2/3}G^2\frac{F^2(0)}{4}J_+^{-1/3} .\label{eqn:e18}\end{equation}
This requires the value of the E-field at $x_o$, or in terms of the boundary layer variables, the value of $F(0)$, where $F$ is the solution of the transformed Painleve equation,
\begin{equation}F''-\frac{1}{2}G^2 F^3 +zF=-1 \label{eqn:e19}\end{equation}
with asymptotic solutions
\begin{eqnarray}
F(z) = \left\{
        \begin{array}{ll}
            -\frac{1}{z} & \quad z\ll 0 \nonumber \\
            \quad \quad \nonumber \\
            \frac{\sqrt{2z}}{G} & \quad z\gg 0
        \end{array}
    \right. \\
\label{eqn:e20}\end{eqnarray}
The solution is obtained numerically using a modified shooting method with the asymptotic solutions as boundary conditions. Initial guesses are supplied by higher-order corrections to the asymptotic expressions for the first few values of $G$. The solution for higher values is obtained by employing a (finite-difference) Taylor expansion of $F(z;G)$ in $\delta G$. $F(z)$ is seen to decrease in the ESC with decreasing selectivity (see Fig.~\ref{fig:f2}). To use this result conveniently in subsequent calculations, the value of $F(0)$ as a function of $G$ is obtained using a fourth-order polynomial interpolation of the results. By evaluating the value of $x_o$ at the limiting salt flux density, it is shown in Fig.~\ref{fig:f3} that decreasing the selectivity increases the ESC thickness and that longer diffusion length to membrane thickness ratios are more sensitive to changes in the selectivity. A more pertinent question is how the selectivity changes couple to the current density itself. This requires use of Eqn.~\ref{eqn:e9} to obtain the dependence of selectivity on applied current. A similar analysis was carried out in~\cite{AbuRjal} up to the limiting current. Here we extend those considerations to the over-limiting case.\\
\indent As stated previously, we impose continuity of the electrochemical potentials, $\mu_{\pm}(x)=\ln{c(x)}\pm\phi(x)$ ,between the ESC at $x=-1$ and the membrane interior as a jump condition across the EDLs. This yields the following expression for the interior membrane boundary concentration at $x=-1$ in the under-limiting regime,
\begin{equation}c_{m,ul} (-1)=\sqrt{\left[1-J_+\left(L-1\right)\right]^2 + N^2/4} \label{eqn:e21}\end{equation}
and in the over-limiting regime, using the concentration $c(x_o)$ of Eqn.~\ref{eqn:e17},
\begin{equation}c_{m,ol}(-1)=\sqrt{\frac{N^2}{4} - \frac{\epsilon^2 J_+}{1+x_o}\left(4G^2-\frac{1}{4}\right)}\label{eqn:e22}\end{equation}
and the usual expression for the concentration at the enriched side,
\begin{equation}c_m(1)=\sqrt{\left[J_+\left(L-1\right)+1\right]^2 + N^2/4} \label{eqn:e23}\end{equation}
To be consistent with the estimate of the ESC width, we have kept the first-order correction term, even though this amounts to a change in the membrane interior concentration of order $O(\epsilon^{\sqrt{4/3}})$. Were we to neglect this contribution (such as the approach taken in~\cite{RZPRE2010}), the qualitative picture would not change.\\
\indent The current-voltage relationship for the over-limiting case can be obtained simply by modifying the under-limiting (LEN) model~\cite{dissertation,AbuRjal},
\begin{eqnarray}V= & \ln{\left[\frac{c_m(1)+ N/2}{c_{m,ul}(-1)+ N/2}\right]} - \left(G+1\right) \ln{\left[\frac{c_r(1)}{c_{\ell}(-1)}\right]} \nonumber \\ & -\frac{GJ_+}{\mathcal{D}}\mathfrak{F}(1) \label{eqn:e24}\end{eqnarray}
The above equation, like the internal membrane concentrations in Eqns.~\ref{eqn:e21} and~\ref{eqn:e22}, is obtained from the continuity of electrochemical potentials. However, upon inspection, it is clear that it may be thought of as a sum of voltage drops across the layers in the system. While it is generally true that one cannot arbitrarily add response coefficients in a multi-layer system~\cite{Kedem}, one can always add voltage drops themselves, provided such drops can actually be resolved in a straightforward way. Thus, one may insert the ESC voltage drop, Eqn.~\ref{eqn:e16} into the sum. It is worth emphasizing that Kirkwood's argument regarding the continuity of electrochemical potentials between phases only applies between the membrane interior and the interior of the ESC and across the EDL in the enriched side. Here, the EDLs are in fact thin, quasi-equilibrium boundary layers separating physically distinct phases. However this is not the case at the artificial boundary, implied by $x_o$. This is a mathematical artifice corresponding to the poles in the asymptotic expansions, and not a Gibbs plane between physically distinct phases. So it is sufficient to demand continuity of concentration, as we did to obtain the estimate of $x_o$. However, it is necessary to modify the concentration at the edge of the quasi-electro-neutral DL and the DL length accordingly. Doing so, one obtains the following approximate form for the current-voltage relationship in the over-limiting regime,
\begin{eqnarray}V= & \ln{\left[\frac{c_m(1)+ N/2}{c_{m,ol}(-1)+ N/2}\right]} - \left(G+1\right) \ln{\left[\frac{c_r(1)}{1-J_+(L
+x_o)}\right]} \nonumber \\ & -\frac{GJ_+}{\mathcal{D}}\mathfrak{F}(1) +\Delta V_{ESC} \label{eqn:e25}.\end{eqnarray}
\section{Results}
\begin{figure}
  \includegraphics[width=3.5 in]{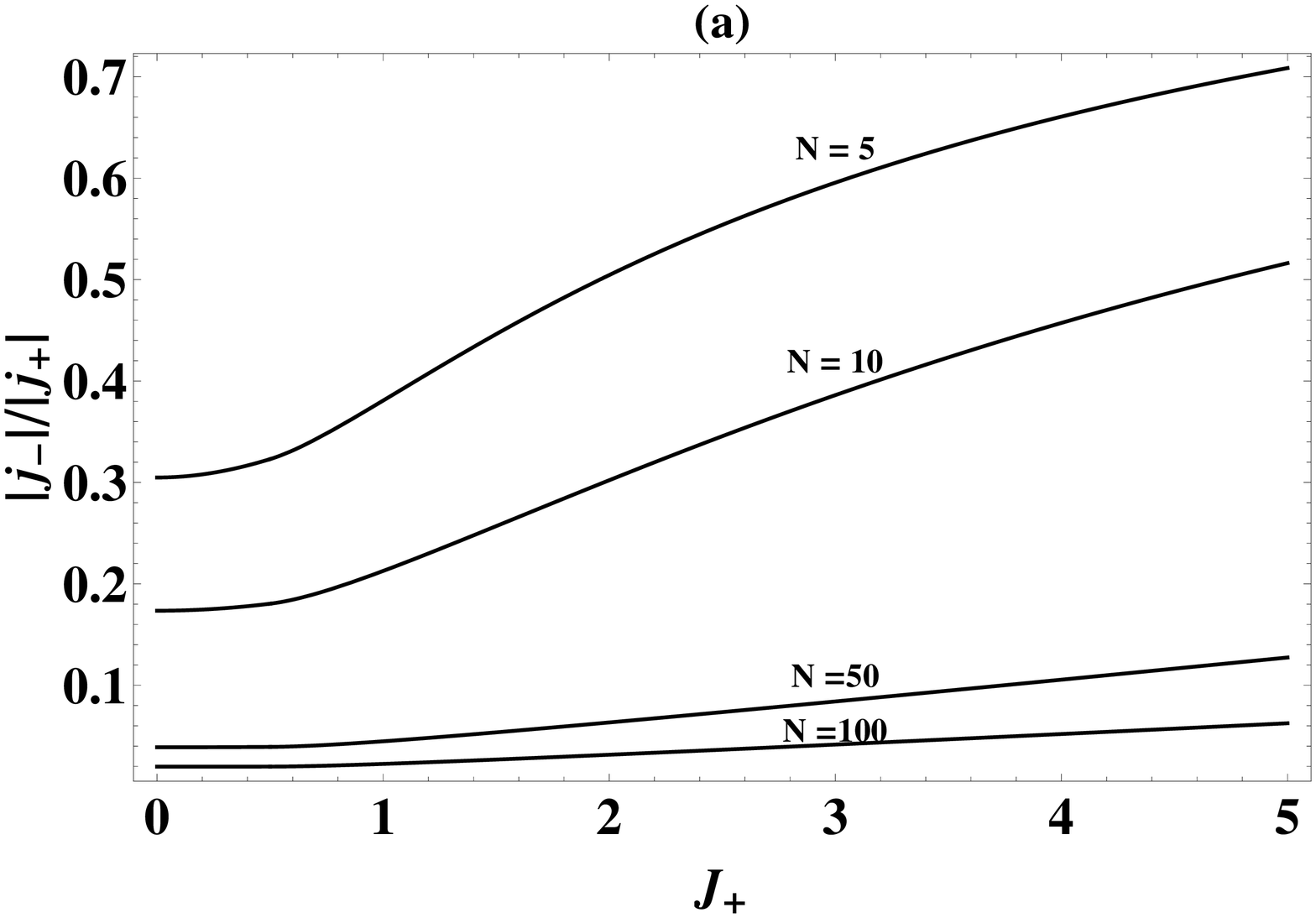}\\
  \includegraphics[width=3.5 in]{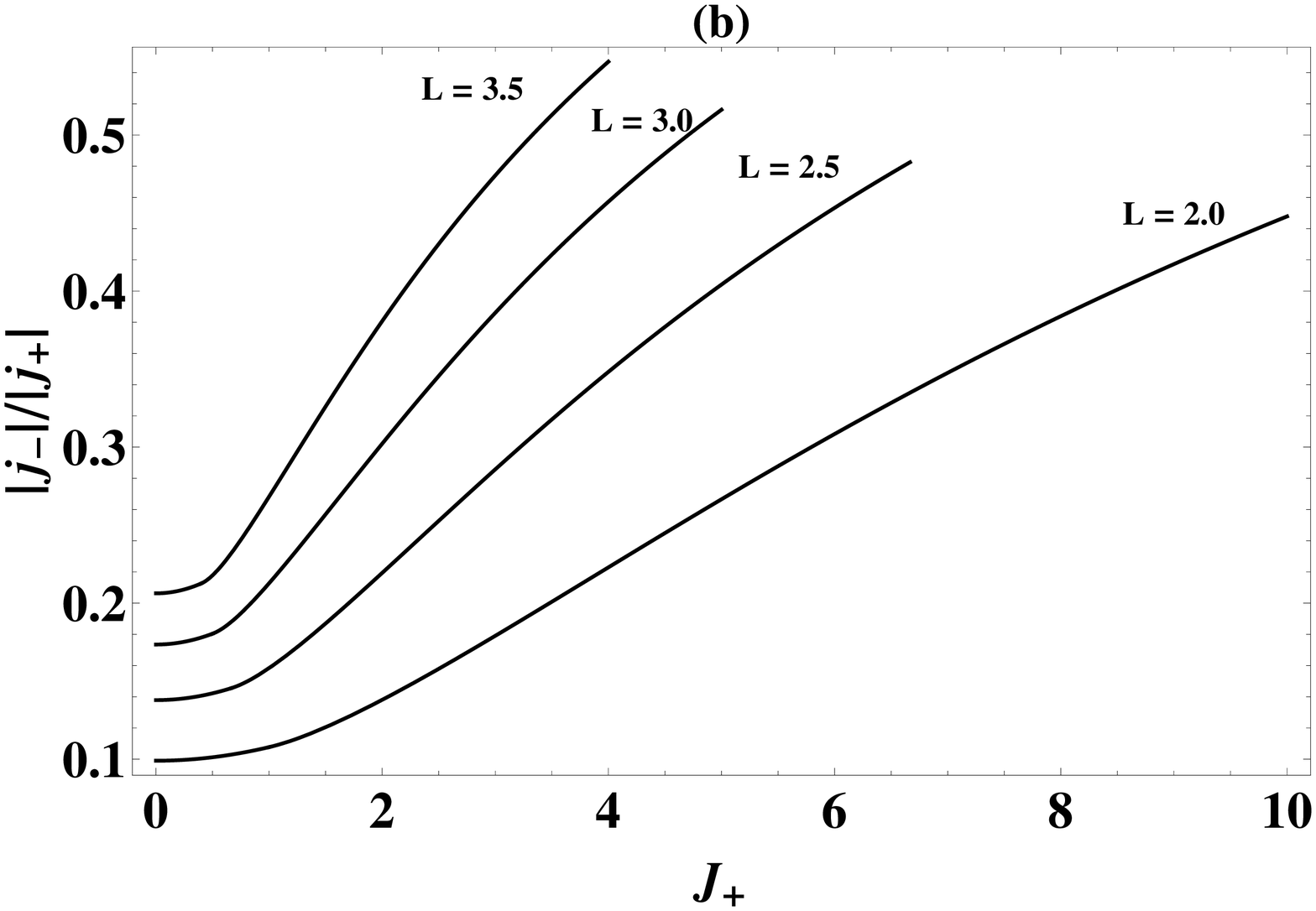}\\
  \includegraphics[width=3.5 in]{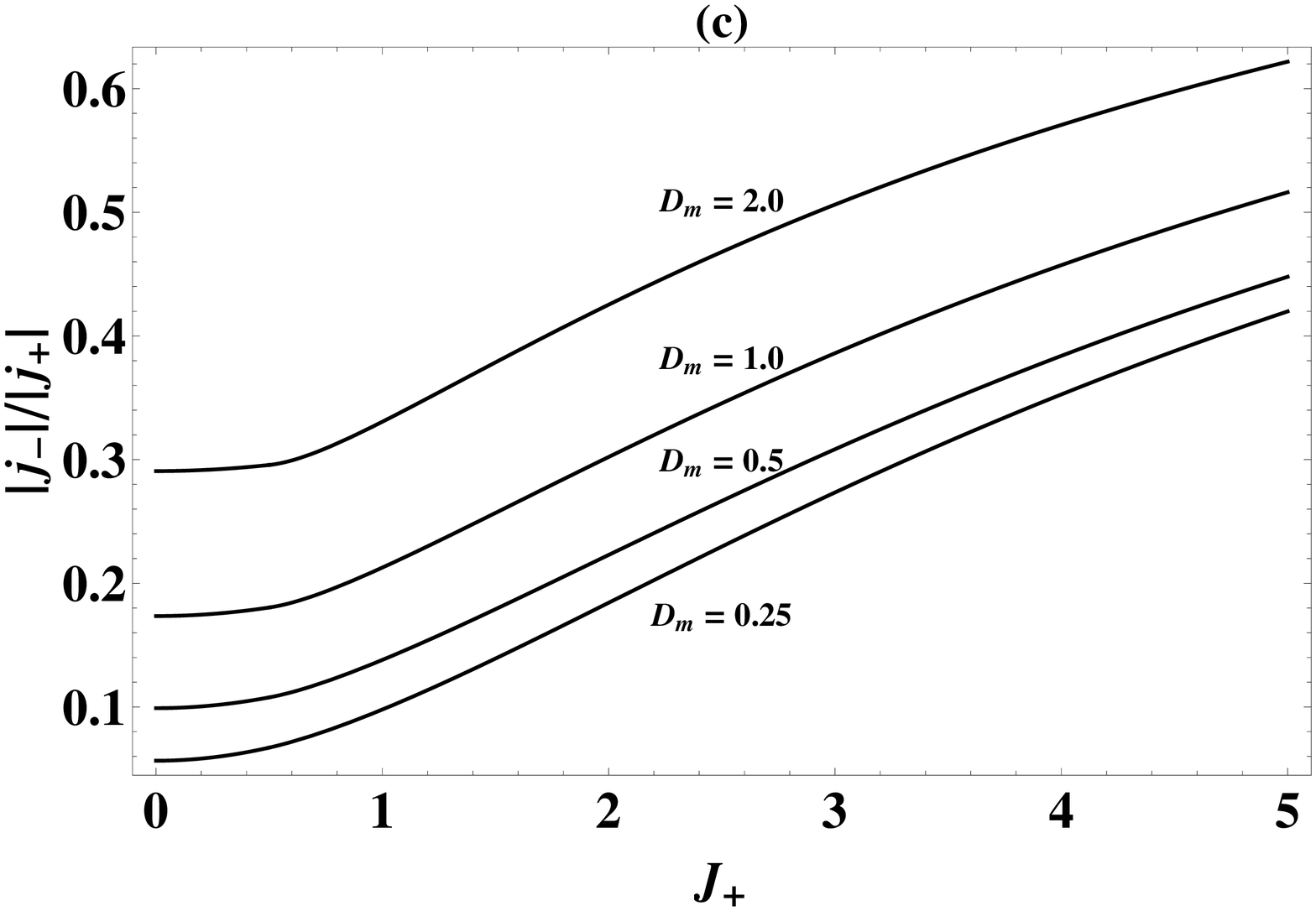}\\
  \caption{Selectivity variation with salt flux density in terms of ratio of co- to counter-ion flux density for various system parameters where the standard case is $N=10$, $L=3$, and $\mathcal{D}_m=1$. Each plot shows a comparison of the standard to other cases; (a) varying fixed charge density $N$, (b) varying diffusion length $L$, and (c) varying membrane diffusivity, $\mathcal{D}_m$. }\label{fig:f4}
\end{figure}

\begin{figure}
  \includegraphics[width=3.5 in]{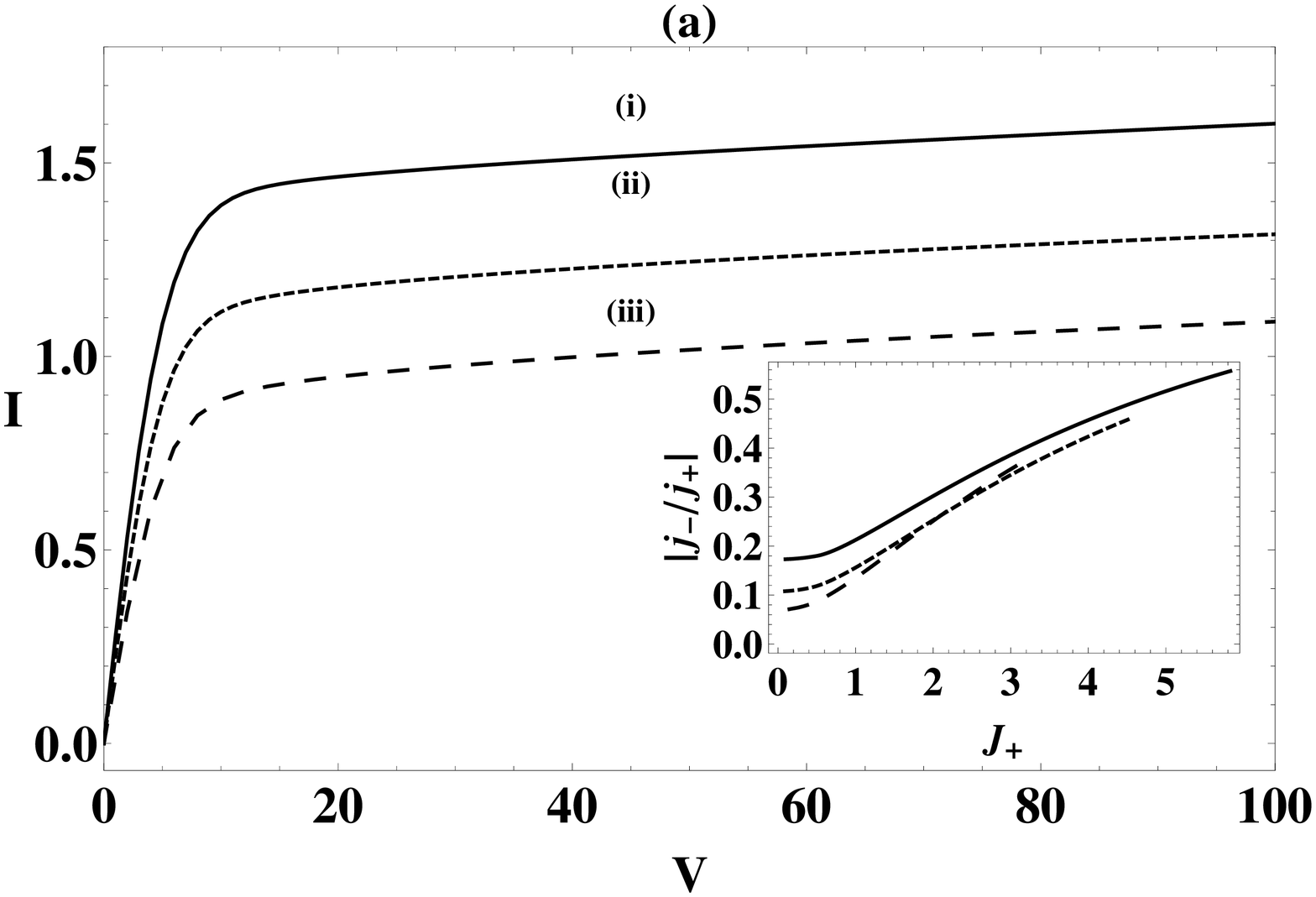}\\
  \includegraphics[width=3.5 in]{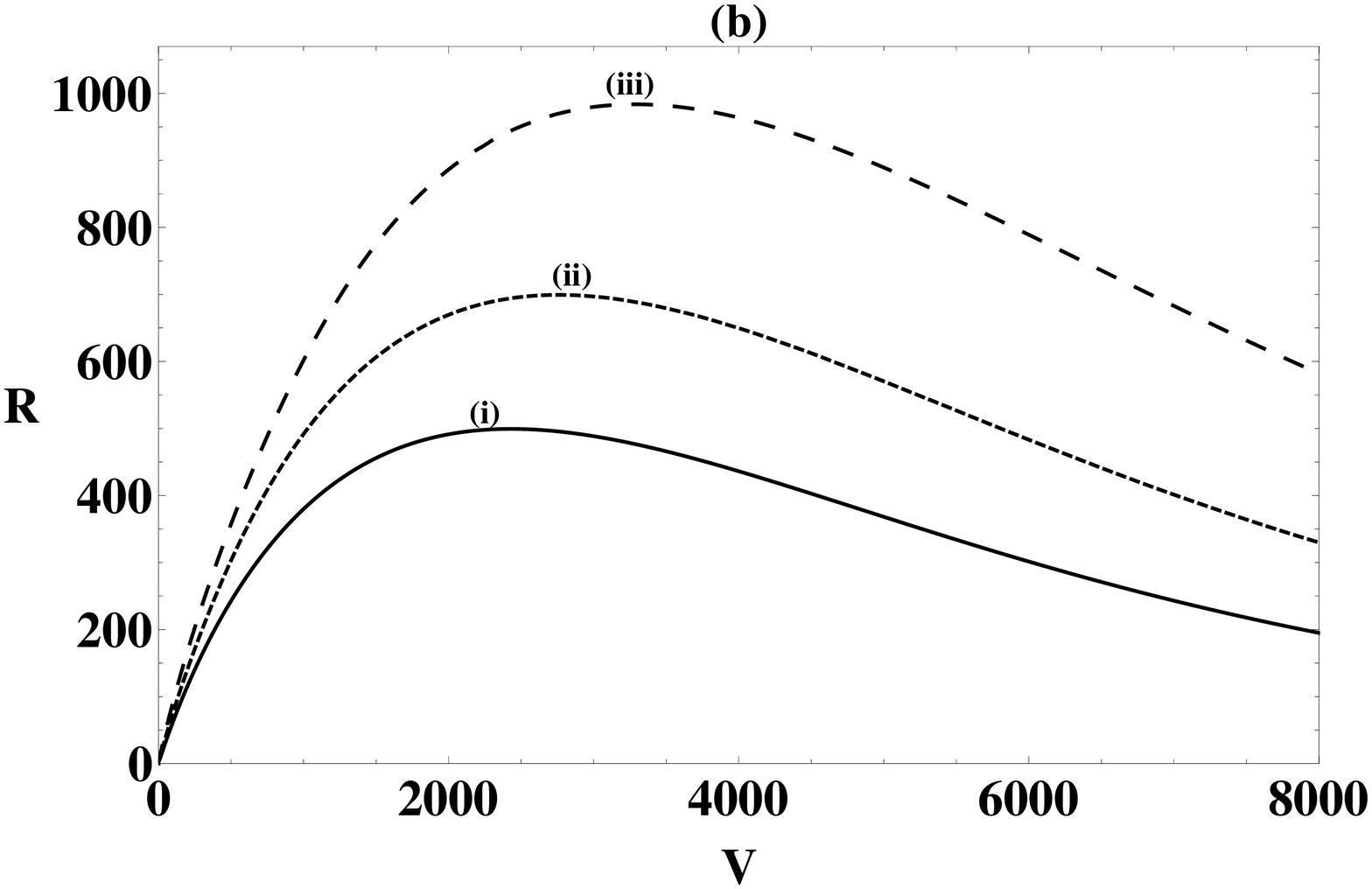}\\
  \caption{Comsol cases $L=2$,$N=10$, bulk diffusivity everywhere and (i) $H=h=1$ corresponding to the standard case of the analytic model, (ii) $H=1$,$h=1/2$, (iii) $H=1$, $h=1/4$. (a) I-V with selectivity ratio inset, (b) R-V}\label{fig:f8}
\end{figure}

\begin{figure}
  \includegraphics[width=3.5 in]{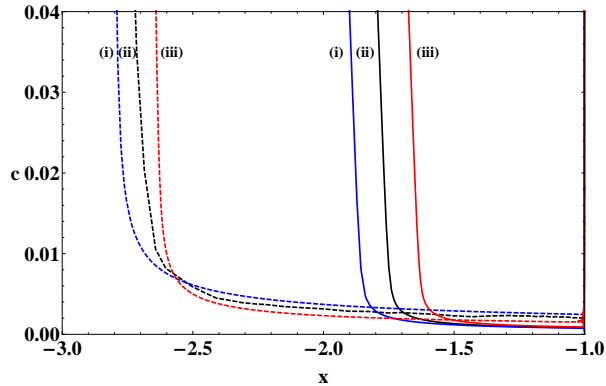}\\
  \caption{Depleted salt concentration profile for numerical 2D case, evaluated on centerline near nanochannel entrance with $L=2$,$N=10$, bulk diffusivity everywhere. Solid lines evaluated for $V=1000$ and dashed lines for $V=7000$ (i) $H=h=1$ corresponding to the standard case of the analytic model, (ii) $H=1$,$h=1/2$, (iii) $H=1$, $h=1/4$.}\label{fig:f9}
\end{figure}

\indent The dependence of the selectivity on salt flux density may be extracted directly from Eqn.~\ref{eqn:e9} by substituting the concentration at $x=-1$, using Eqn.~\ref{eqn:e21} or~\ref{eqn:e22} and that at $x=1$ Eqn.~\ref{eqn:e23}, for the appropriate regime. This is shown in Fig.~\ref{fig:f4}. The ratio of the absolute value of co- to counter-ion flux density is plotted against overall salt flux density for a range of parameters, with an increase in the ratio corresponding to loss of selectivity. The range of salt flux density is the same for all three cases, from zero to five times the limiting value. The selectivity decreases as the salt-flux is increased because the difference in co- and counter-ion electrochemical potential across the charge-selective region increases. This corresponds to both an increase in conductivity and increasing co-ion concentration within the charge-selective region. In agreement with previous studies~\cite{AbuRjal}, the change in the under-limiting regime is rather modest. However, since the enriched concentration has no imposed upper bound, this difference can be expected to increase as $J_+$ exceeds $J_{+,lim}$.\\
\indent In Fig.~\ref{fig:f4}a, the change in selectivity is shown for several values of the fixed charge density. Unsurprisingly, as the charge density increases--corresponding to increased restriction of the co-ion flux--both the overall selectivity and change in selectivity decrease. The cases in Figs~\ref{fig:f4}b and c bear some relationship to the effects of field-focusing in the sense that increased field-focusing corresponds to both increasing flux density in the focused region and effective shortening of the diffusion length~\cite{YCffPRE}. For different diffusion lengths $L$, the total change in selectivity across the range of salt flux density, from $J_+=0$ to $J_+=5 J_{+,lim}$, is the same for all $L$. However, the overall selectivity itself decreases with increasing $L$ because the concentration at the enriched boundary is higher for the same salt flux density. The same is essentially true of varying $\mathcal{D}_m$; the total change in selectivity is independent of the effective diffusivity in the charge-selective region. However, while the limiting salt flux density in the diffusion layers is the same for all cases, a decrease in effective diffusivity may be thought of as an increase in effective salt flux density inside the charge-selective region. So a comparably small change in concentration at the enriched side, corresponding to a comparably small total change in electrochemical potential across the membrane, leads to lower overall selectivity without shifting the voltage.\\
\begin{figure}
  \includegraphics[width=3.5 in]{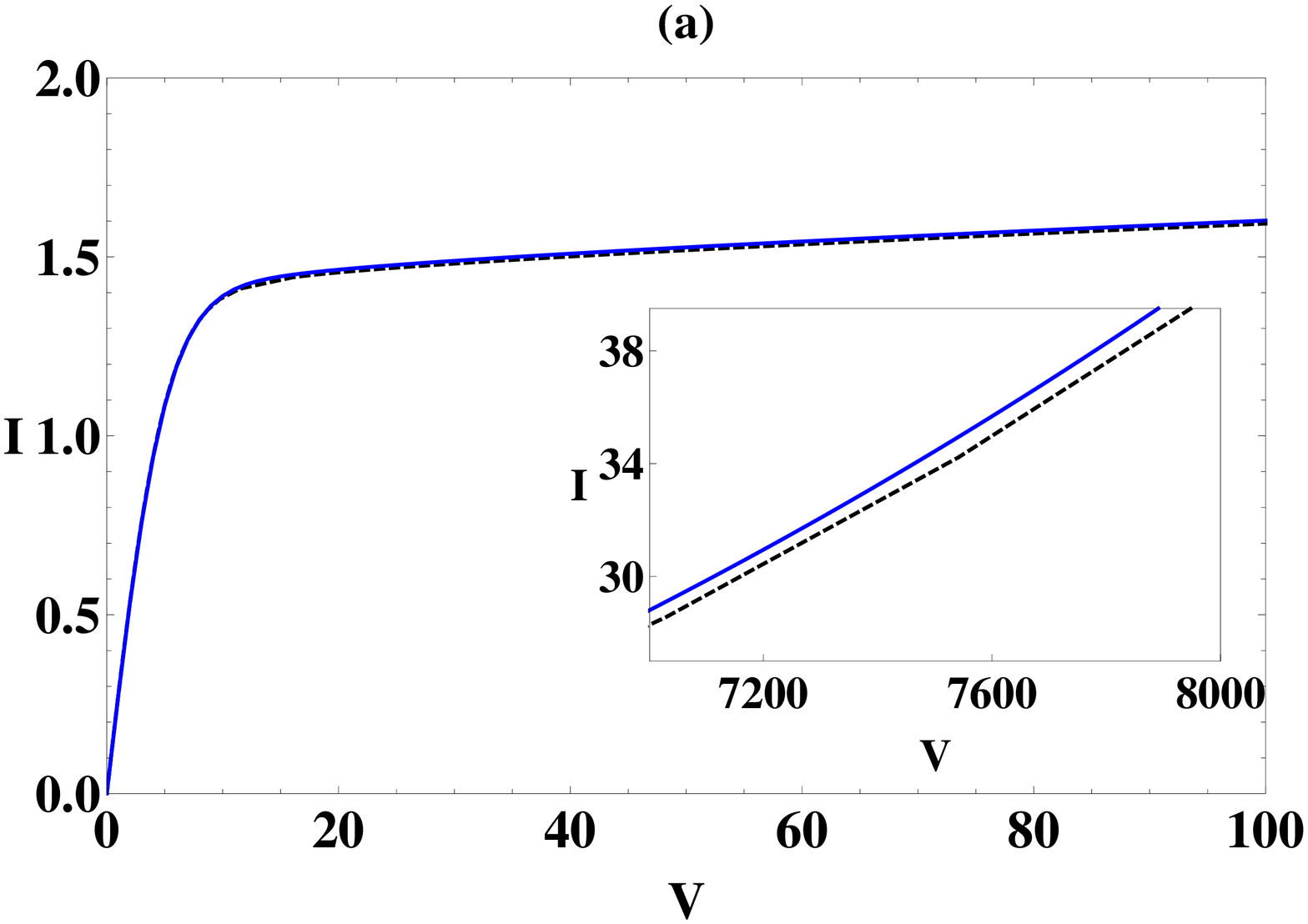}\\
  \includegraphics[width=3.5 in]{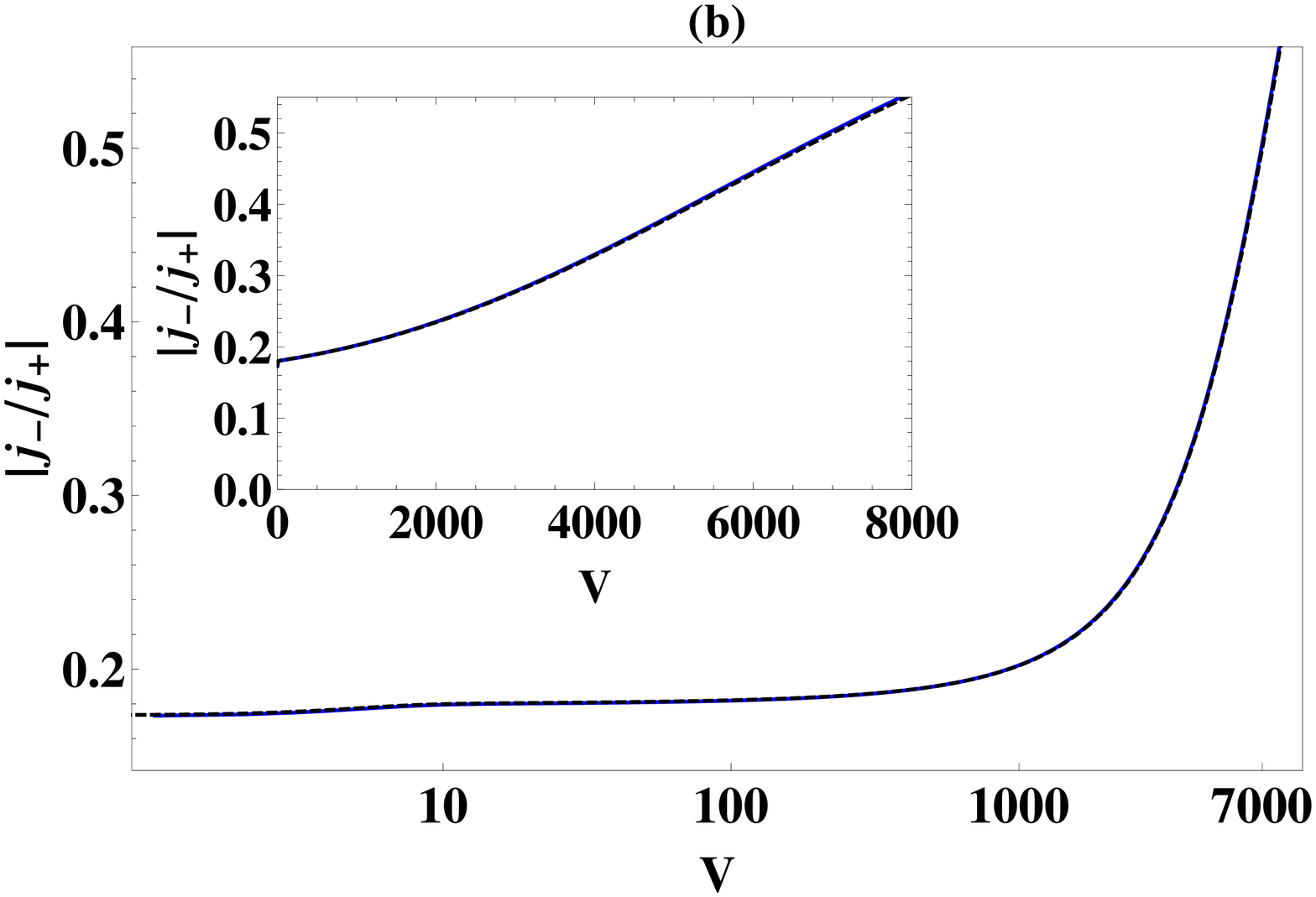}\\
  \caption{Comparison of theoretical model (dashed black) to COMSOL calculations (solid blue) for a 1D PNP model, both using the standard parameter set $N=10$,$L=3$, and $\mathcal{D}_m=1$. (a) Current-voltage response is shown in the top figure and changes in selectivity in the bottom (b). Inset in top figure shows discrepancy between approximate model and COMSOL grows with voltage.}\label{fig:f5}
\end{figure}
\indent To validate our approach, we compare the results of the above model with numerical calculations made using COMSOL software, shown in Fig.~\ref{fig:f5}. The agreement between the approximate analytic model and the numerical model is quite good, diverging appreciably only at the highest applied voltages where the analytic model under-predicts the current density by a few percent. The two models yield essentially identical selectivity response across the entire voltage range, showing that the bulk of the selectivity change occurs at higher voltages.\\
\indent Concerning the resistance maximum, both the maximum value of the resistance and the voltage at which it occurs are related to the changes in selectivity, and thus tied to the other parameters in the system. The change in selectivity itself does not require an accurate value of the concentration in the ESC; similar change would be obtained by simply equating the over-limiting concentration to zero. However, the very existence of the maximum itself arises from the fact that not only is the concentration non-zero, but increases with current beyond the limiting value (see Eqn.~\ref{eqn:e17}). The basic mechanism behind the maximum may be understood as follows: Initially, the increase in the extent of the depleted region, as indicated by $x_o$, causes a rapid increase in the ESC resistance. However, as the ESC concentration increases, the effective conductivity of the depleted region increases both within the ESC and at the edge of the quasi-electro-neutral region, giving rise to the maximum.\\
\begin{figure}
  \includegraphics[width=3.5 in]{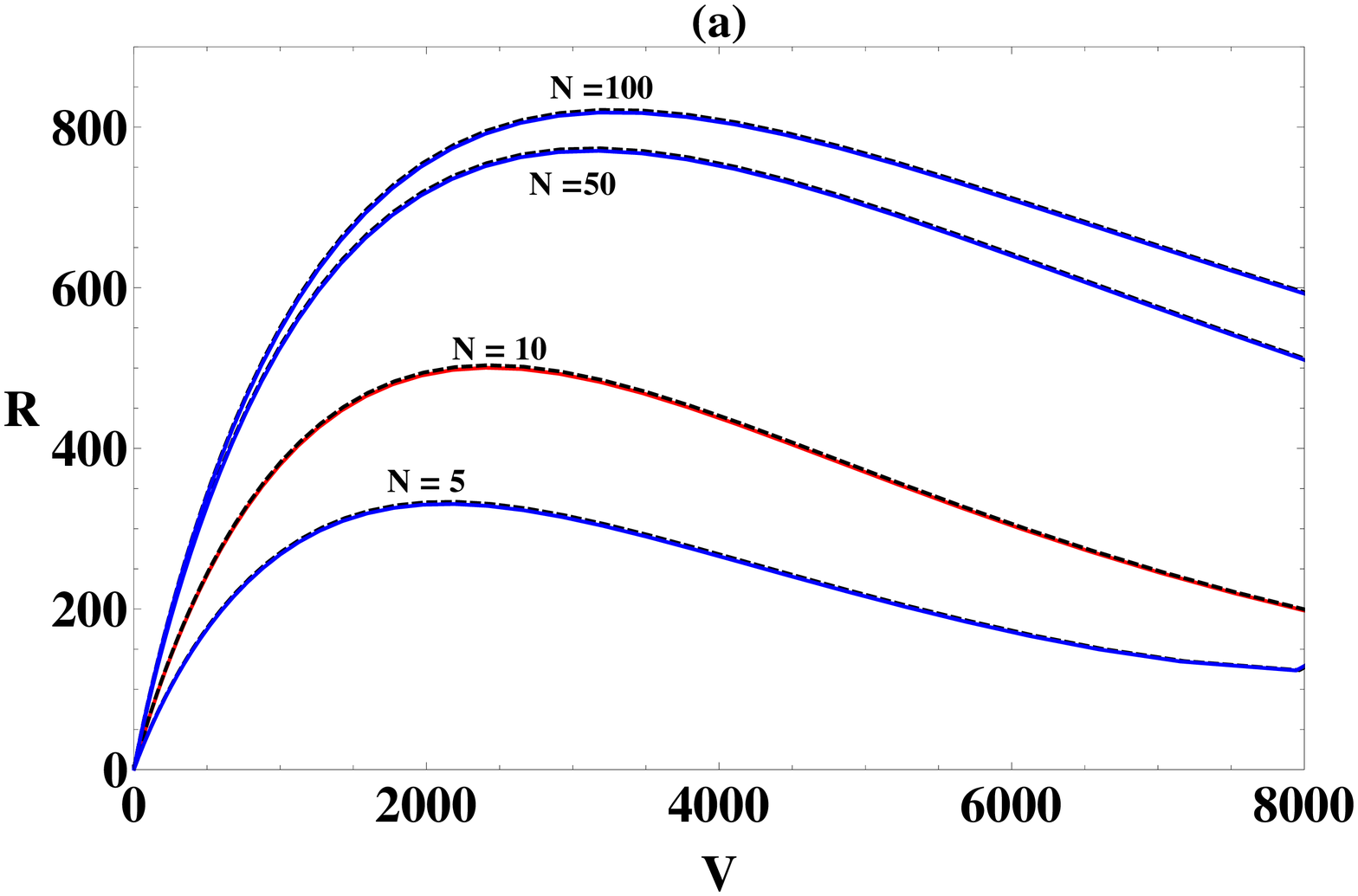}\\
  \includegraphics[width=3.5 in]{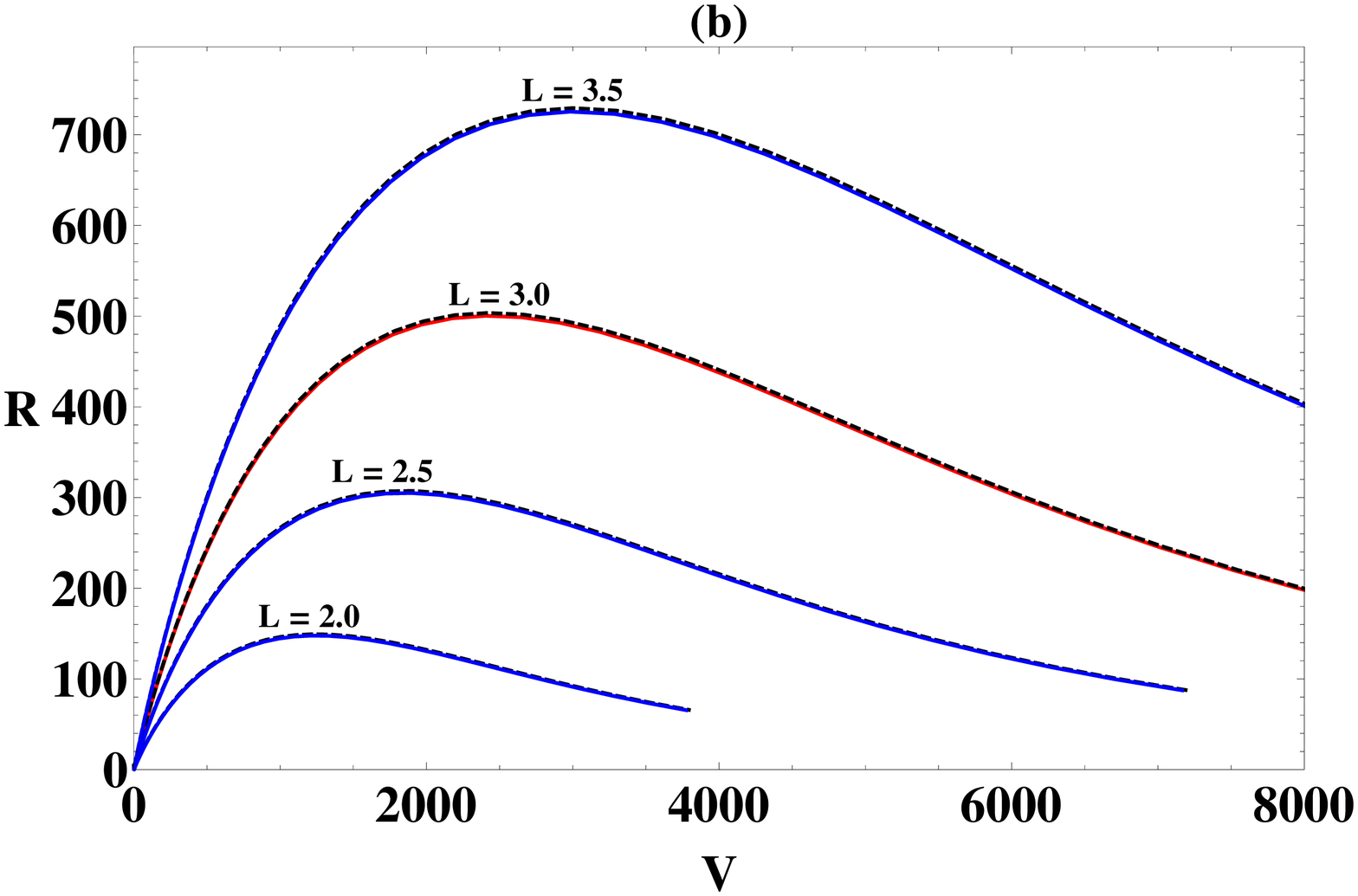}\\
  \includegraphics[width=3.5 in]{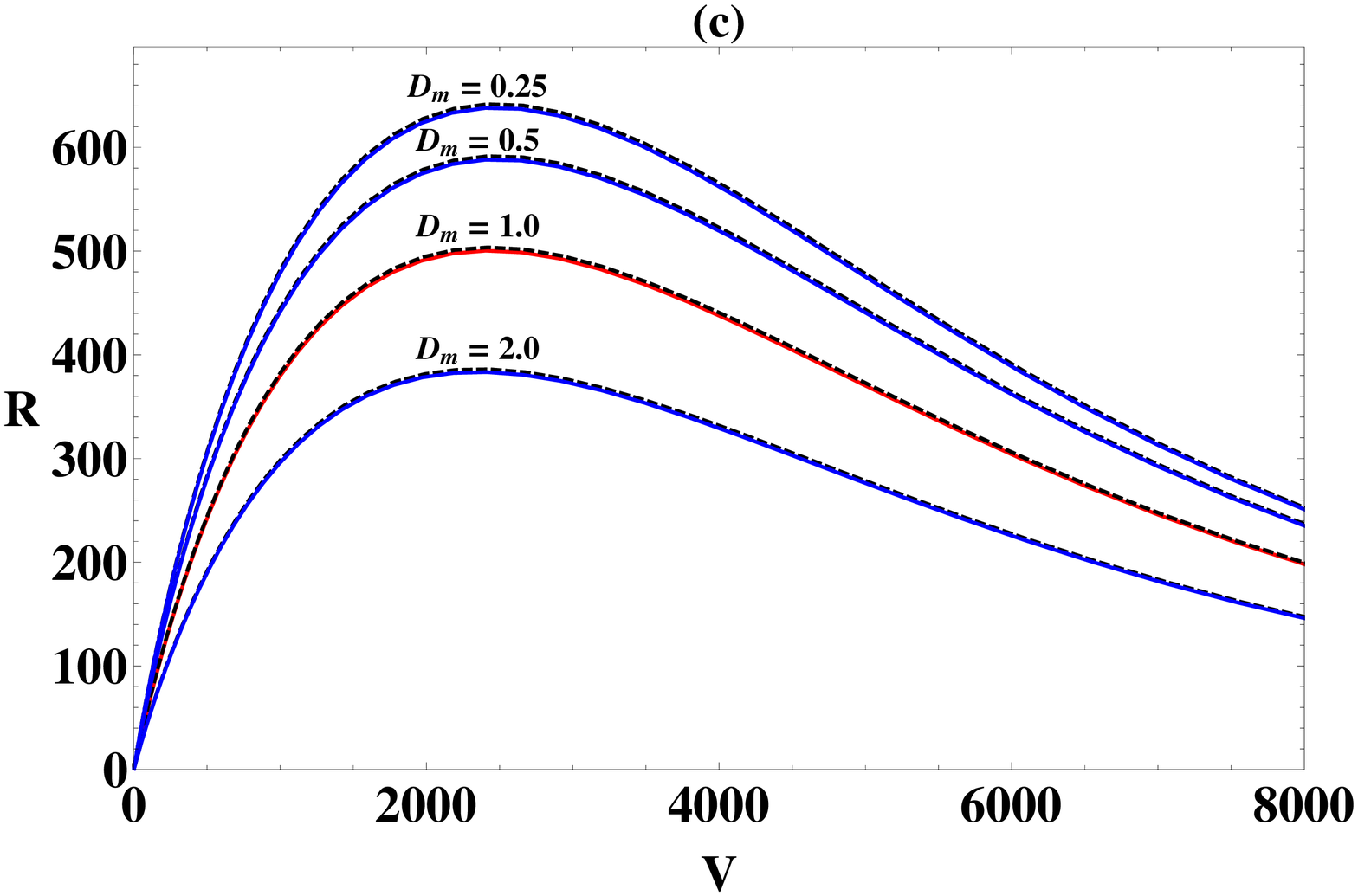}\\
  \caption{Resistance vs. voltage curves, for cases matching figure 4. The `standard' case (solid red) has $N=10$, $L=3$, and $\mathcal{D}_m$. The total resistance (dashed black) and ESC resistance (solid blue or red) are shown for each case.}\label{fig:f6}
\end{figure}
\indent By considering the variation of the concentration $c(x_o)$ (Eqn.~\ref{eqn:e17}) along with selectivity variation one can understand the results shown in Fig.~\ref{fig:f6}. The parameters correspond to the same cases as shown in Fig.~\ref{fig:f4}. Both total resistance and ESC resistance are shown for each case. For the range of parameters considered here, the resistance is almost totally dominated by the ESC resistance.\\
\indent Fig.~\ref{fig:f6}a shows the resistance-voltage curves for different values of the fixed charge density, $N$. For large values of $N$, the selectivity is high and, as seen in Fig.~\ref{fig:f4}a, changes little with applied voltage. The co-ion concentration inside the charge-selective region remains low and the counter-ion concentration is governed primarily by the value of $N$ itself. Thus $G$ remains very close to the ideal value, $G=1$ for all voltages and the ESC concentration increases slowly with voltage, being governed primarily by the slow increase of $J_+^{2/3}$. So the overall resistance is high and the slow change in concentration imply a maximum resistance at higher voltage than the other cases. For smaller values of $N$, the change in $G$ (equivalently, the ratio $|j_-/j_+|$) is larger, resulting in increasing $G$ faster than the decrease in $F(0)$. Hence there is a larger increase in the ESC concentration and lower resistance occurring at a lower voltage for smaller $N$.\\
\indent The dependence of the resistance-voltage curve on diffusion length $L$ is more indicative of the changes in the limiting salt flux density and in the enriched concentration than the increased resistance of the longer diffusion layers. This can be expected to differ from a true 2D/3D case, where field-focusing plays a strong role~\cite{YCffPRE}, or a situation in which non-uniform fluid flow can alter the depleted concentration profile~\cite{Yaroschuk,NBruus}. The maximum occurs at higher voltages for longer diffusion lengths because the total change in selectivity occurs over a smaller range of salt flux density while the actual value of the salt flux density is lower for longer $L$ (see the discussion associated with Fig.~\ref{fig:f4}), corresponding to a smaller increase in ESC concentration for a given change in total applied voltage. Thus it takes a larger overall voltage to reach the resistance maximum. The reasoning for changing diffusivity is somewhat similar in the sense that higher effective salt flux in the membrane for smaller values of $\mathcal{D}_m$ corresponds to an overall higher selectivity, hence lower co-ion concentration and flux. However, because it does not correspond to an actual modification of the concentrations or electric potentials adjacent to the membrane, the voltage does not shift.\\
\begin{figure}
  \includegraphics[width=3.0 in]{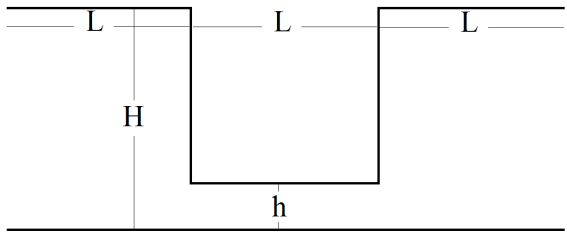}\\
  \caption{COMSOL geometry sketch, note that $L$ here does not correspond directly to that in the 1D model.}\label{fig:f7}
\end{figure}
\indent To gain some insight into the influence of field-focusing on the structure of the ESC, selectivity variation, and dc response, a 2D numerical solution for the PNP equations is obtained using COMSOL. The geometry is shown in Fig.~\ref{fig:f7}. A uniform volumetric charge density, $N=10$, is specified in the constricted `nanochannel' region. No surface charge is specified on the channel walls, so that we may study the effects of field-focusing on the ESC in isolation from surface-conduction. Fig.~\ref{fig:f8}a shows the current-voltage characteristics in the main figure, which are qualitatively similar to those obtained for either decreasing diffusion-length $L$ or decreasing $\mathcal{D}_m$ in the 1D model. However, note the selectivity response (inset) shows that the increased field-focusing results in both higher selectivity initially, and a greater overall change in selectivity with increasing salt flux (equivalently, voltage.) It is important to keep in mind that the numerical calculations are done under potentiostatic conditions, as opposed to the galvanostatic conditions employed in analytic model. Therefore, the voltage range is kept fixed and the salt-flux and current densities are determined in response to the voltage. Thus the same voltage range produces a change in selectivity over a smaller range of salt-flux density for an increasing degree of field-focusing. The resistance-voltage curve in Fig.~\ref{fig:f8}b further illustrates the difference. While the value of the resistance increases, in this case corresponding to a smaller nanochannel height for a fixed microchannel height, the resistance maximum shifts to a higher voltage. This stands in contrast to both decreasing $\mathcal{D}_m$ and decreasing $L$, where by analogy we would expect the increased degree of field focusing to correspond to either no shift in the voltage of the maximum, or a shift to lower voltages.\\
\indent To better understand what is happening in the 2D case, we consider the evolution of the 2D ESC with voltage. The centerline salt concentration near the nanochannel entrance is plotted in Fig.~\ref{fig:f9} for voltages below and above the maximum. Note that, as one might expect, increasing field-focusing corresponds to a smaller ESC at the same voltage. However, the ESC concentration undergoes less of an increase with voltage when the degree of field-focusing is higher. While this seems at first counter-intuitive, it is worth noting that the range in salt flux density for the same voltage range is reduced considerably. Thus, at a given voltage, a lower salt flux density drives the change in ESC salt concentration leading to a slower increase in ESC concentration with increased field focusing.\\
\section{Conclusions}
\indent We have presented theoretical and numerical results concerning the connection between variation of selectivity, ESC structure, and dc over-limiting resistance of a 3-layer model electro-diffusive system with a non-ideal charge-selective element. This approach extends previous studies concerning the change in selectivity for a similar system in the under-limiting regime~\cite{AbuRjal}. More importantly, the present study represents progress towards developing a model capable of capturing the OLC dc response of a fabricated micro-nanochannel system. The majority of previous work has relied on models which assumed ideal permselective transport, which make it difficult to understand the relationship and interaction of the ESC with other OLC mechanisms such as surface conduction~\cite{ManiB,KimPRL2015} or non-uniform fluid flow~\cite{Yaroschuk,NBruus}. It was recently shown that non-ideal selectivity can lead to a bulk electro-convective instability within the locally electro-neutral approximation~\cite{RZnonidealPRL}, and the present work serves to demonstrate that the structure of the ESC itself is affected by both non-ideality and field-focusing. It is reasonable to assume such structural change can affect electro-convective stability as well, and inclusion of these effects could be expected to enrich the present understanding of electro-convection in OLC and high-current transport in experimental micro-nanochannel devices.

\begin{acknowledgments}
We wish to acknowledge Israel Science Foundation, grant number 2015240, the Stephen and Nancy Grand Water Research Institute, grant number 2017720, for financial support.\\
\end{acknowledgments}

\renewcommand{\refname}{References}

\end{document}